\documentstyle[11pt,newpasp,twoside,epsf]{article}
\def\edcomment#1{\iffalse\marginpar{\raggedright\sl#1\/}\else\relax\fi}
\reversemarginpar

\begin{document}
\title{Properties of Dark Matter Halos in Disk Galaxies}
 \author{Roelof S. de Jong}
%\affil{Space Telescope Science Institute, 3700 San Martin Dr., Baltimore, MD 21218, U.S.A.}
\affil{STScI, 3700 San Martin Dr., Baltimore, MD 21218, U.S.A.}
\author{Susan Kassin}
\affil{Department of Astronomy, Ohio State University, 140 West 18th Avenue, Columbus, OH 43210-1173, U.S.A.}
\author{Eric F. Bell}
%\affil{Max-Planck-Institut f\"ur Astronomie, K\"onigstuhl 17, D-69117 Heidelberg, Germany}
\affil{MPIA, K\"onigstuhl 17, D-69117 Heidelberg, Germany}
\author{St\'ephane Courteau}
\affil{Department of Physics and Astronomy, University of British Columbia, 6224 Agricultural Road, Vancouver, BC V6T 1Z1, Canada}

\begin{abstract}

We present a simple technique to estimate mass-to-light ($M/L$) ratios
of stellar populations based on two broadband photometry measurements,
i.e.\ a color-$M/L$ relation. We apply the color-$M/L$ relation to
galaxy rotation curves, using a large set of galaxies that span a
great range in Hubble type, luminosity and scale size and that have
accurately measured H{\small I} and/or H$\alpha$ rotation
curves. Using the color-$M/L$ relation, we construct stellar mass
models of the galaxies and derive the dark matter contribution to the
rotation curves.

We compare our dark matter rotation curves with adiabatically
contracted Navarro, Frenk, \& White (1997, NFW hereafter) dark matter
halos. We find that before adiabatic contraction most high surface
brightness galaxies and some low surface brightness galaxies are well
fit by a NFW dark matter profile. However, after adiabatic
contraction, most galaxies are poorly fit in the central few kpc. The
observed angular momentum distribution in the baryonic component is
poorly matched by $\Lambda$CDM model predictions, indicating that the
angular momentum distribution is not conserved during the galaxy
assembly process.  We find that in most galaxies the dark matter
distribution can be derived by scaling up the H{\small I} gas
contribution. However, we find no consistent value for the scaling
factor among all the galaxies.

\end{abstract}

%\section{Introduction}

For a considerable time galaxy rotation curves have been providing
some of the strongest evidence for Dark Matter (DM) in the Universe
(e.g., Freeman 1970; Rubin 1978; Bosma 1981).  A long standing problem
has been the derivation of the DM mass distribution from the observed
rotation curves. Interestingly, this problem is not due to the unknown
nature of the DM component, but is mainly due to the unknown mass
scaling of the observed stellar component. This has often been solved
by using the maximum disk solution, i.e.\ assuming the $M/L$ ratio of
the stellar component does not change with radius and scaling it up by
the maximum amount allowed by the rotation curve (e.g., van Albada et
al.~1985). While this is perfectly legitimate to prove the presence of
DM, once we accept the existence of DM, there is little reason to
believe that all disks are close to maximum (Courteau \& Rix 1999; but
see Sancisi, these proceedings). Especially in light of currently
popular galaxy formation scenarios based on Cold Dark Matter (CDM)
cosmologies, we would not expect that most galaxies are close to
maximum disk. Here we explore what the DM scaling relations are once
we constrain the stellar mass components by stellar population
modeling.

%?- comparison angular momentum (relevant gal formation models)
%?- HI scaling

\section{Stellar Population Mass-to-Light Ratios}

To constrain the stellar mass distribution in our sample galaxies, we
use the color-$M/L$ relationships for mixed stellar populations of
Bell \& de Jong (2001). In this paper we showed that for almost any
reasonable star formation history and consistent chemical evolution of
a galaxy, the stellar population models predict a strong correlation
between the optical color of the population and its $M/L$. This was
shown to be robust against gas inflow and outflow, different star
formation prescriptions and mild starbursts (Fig.\,1a).
% and is mainly due to degeneracies in the optical
%color-$M/L$ plane for most ages and metalicities of simple stellar
%populations and to some extend even dust reddening.

\begin{figure}
%\fbox{\plotfiddle{salp_mass_burste.ps}{3cm}{0}{90}{90}{-250}{-350}}
%\plotfiddle{maxdisk.ps}{4cm}{0}{1}{1}{10}{10}
%\plottwo{salp_mass_burste.ps}{maxdisk.ps}}
\mbox{
\epsfysize=5.8cm
\epsfbox[74 355 216 513]{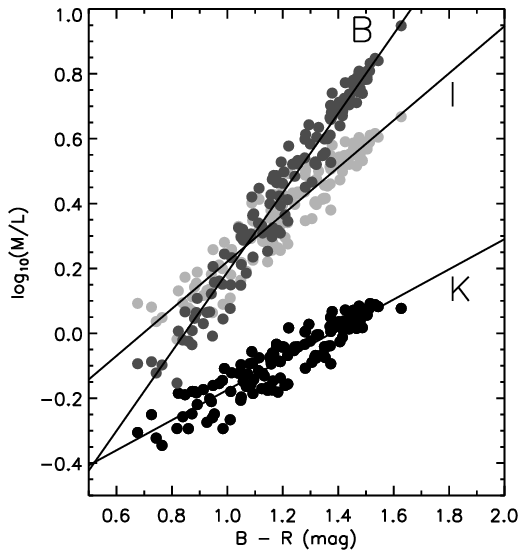}
\epsfysize=5.8cm
\hspace{1.cm}
\epsfbox[188 370 410 597]{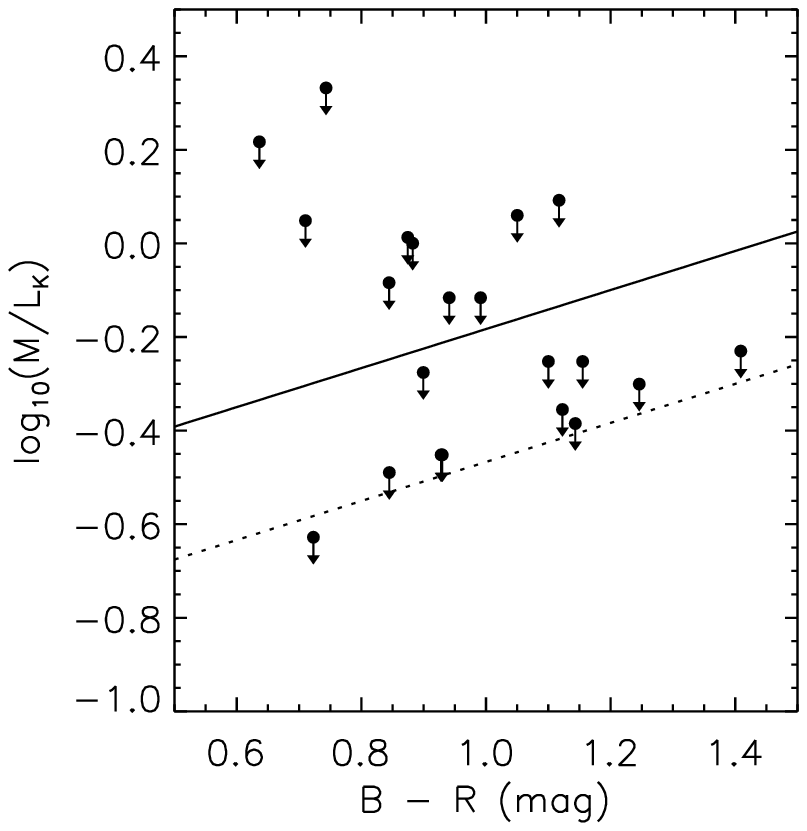}
}
\caption{
{\bf Left (a):} 
Bell \& de Jong (2001) model stellar population $M/L$ ratios in the $B$, $I$ and $K$-band
versus $B$--$R$ color (points) and the fitted slopes (lines). 
{\bf Right (b):} The maximum disk $M/L_K$ ratios for the Verheijen (1997)
galaxies (upper limit symbols) versus $B$--$R$ color. The solid line
shows the stellar population model predictions for a Salpeter IMF
(solid line) and a renormalized Salpeter IMF (dotted line).}
\end{figure}

The one major assumption we had to make was that all galaxies have
similar stellar Initial Mass Functions (IMFs), independent of their
other properties. The IMF used in our color-$M/L$ analysis determines
the slope of the relation, but not the zero-point. This reflects the
fact that the high mass end of the IMF determines the luminosity and
the color of a population, but the low mass end sets its mass,
shifting the relation in Fig.\,1a up and down.

%Therefore, we can add and subtract stars at the low mass end of
%the IMF without changing the population color or luminosity, hence shifting the relation in %Fig.\,1a up and down.

We used the maximum disk $M/L_K$ limits of the Ursa Major galaxies
observed by Verheijen (1997) to constrain the IMF. In Fig.\,1b we show
the maximum allowed $M/L_K$ ratios for these galaxies (hence the upper
limit symbols) versus their $B$--$R$ color. We compare these to the
predictions of our models for a standard Salpeter IMF (solid diagonal
line). If our models are correct, all $M/L$ ratios of galaxies should
scatter around this line. Clearly, this is in conflict with
observations, as some galaxy upper limits are below the line, and thus
using the $M/L$ of our model would over-predict the rotation curve of
these galaxies. Hence, we use a normalization that is lower by 0.3 dex
(dotted line). In principle, the normalization could be much lower,
but there are many indications that in particular high surface
brightness galaxies are close to maximum disk (see e.g.\ Weiner, these
proceedings) and therefore we chose a normalization that has these
galaxies close to maximum disk. If our color-$M/L$ relation is correct
and all true $M/L$ values are near this line, clearly many galaxies
are substantially sub-maximal, with some galaxies in the top part of
the diagram being sub-maximal by a factor of 10. We use here the
updated color-$M/L$ relations of Bell et al.~(2003), which were
calibrated with SDSS and 2MASS data.

\section{Rotation Curve Modeling}

%\section{The Data}

We are currently collecting data for a large sample of galaxies with
high quality rotation curves with a broad range in galaxy structural
parameters (luminosity, scale size, etc.). For each galaxy we aim to
have optical and near-IR surface photometry, H{\small I} synthesis
observations of column densities and rotation curves and high
resolution inner rotation curves from H$\alpha$ or CO measurements.
We expect a final sample of over 100 galaxies; results for 25 are
presented here.

\begin{figure}[t]
%\fbox{\plotfiddle{salp_mass_burste.ps}{3cm}{0}{90}{90}{-250}{-350}}
%\plotfiddle{maxdisk.ps}{4cm}{0}{1}{1}{10}{10}
\epsfysize=8.cm
\epsfbox[-120 161 511 651]{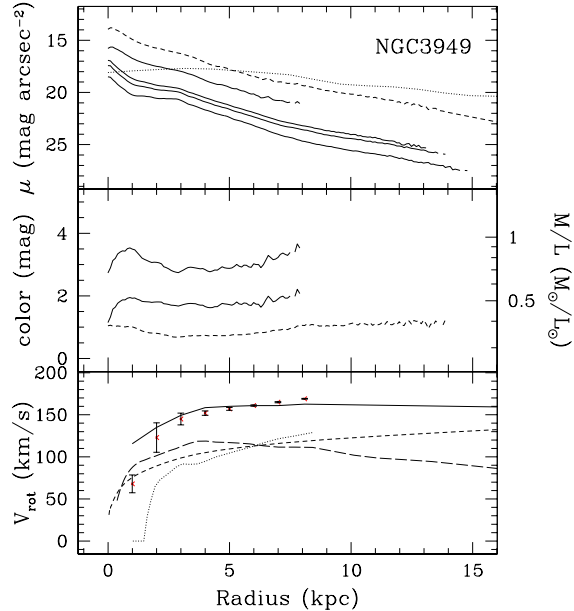}
\caption{({\bf top}) The observed surface brightness profiles of
NGC\,3949 in $K$, $R$, $V$, and $B$-band (solid lines). The derived
surface density in stellar mass (dashed) and H{\small I} (dotted) are
shown on an arbitrary normalization. ({\bf middle}) The $B$--$K$,
$I$--$K$ and $B$--$R$ color profiles, where the dotted  $B$--$R$
profile has been calibrated in terms of $M/L_K$ on the right hand
side. ({\bf bottom}) The observed rotation curve is indicated by crosses
with error bars. We also show the calculated baryonic rotation curve
(long dashed line), the derived DM rotation curve by subtracting the
baryonic curve from the observed one in quadrature (dotted line), the
best fit adiabatically contracted NFW profile (short dashed line) and
the sum of baryonic and NFW rotation curves (solid line).}
\end{figure}

Fig.\,2 shows our method to determine DM distributions in our
galaxies. Using the color profiles of the galaxies, we determine the
radial $M/L$ profiles using the color-$M/L$ relations (middle
panel). This allows us to turn the surface brightness profiles into
surface mass density profiles (top panel). This not only lets us take
the radial variation in $M/L$ into account, but more importantly, it
provides us with a consistent way of scaling stellar $M/L$ ratios from
one galaxy to the next. We add the neutral gas component by scaling
the observed H{\small I} surface density by 1.32 to account for He and
calculate the resulting baryonic rotation curve (bottom panel). As
shown, one can now calculate the DM rotation curve by quadratically
subtracting the baryonic curve from the observed curve, but this
generally yields noisy DM curves in the region dominated by the stars.

\section{Dark Matter Halo Scaling Relations}

In order to avoid the problem of noisy DM curves, one often fits
parametrized DM halo models. Here we show results of fits using Navarro,
Frenk, \& White (1997, NFW) DM halos, adiabatically contracted by the
calculated baryonic component according to the formalism of Blumenthal et
al.~(1986) (see Dutton et al.~2003 for details). While we find that
most high surface brightness galaxies and many low surface brightness
galaxies are fitted by NFW DM profiles before adiabatic contraction,
after contraction the model rotation curves of most galaxies
over-predict the observed rotation in the central regions. 

\begin{figure}
\mbox{
\epsfysize=4.5cm
\epsfbox[44 169 528 539]{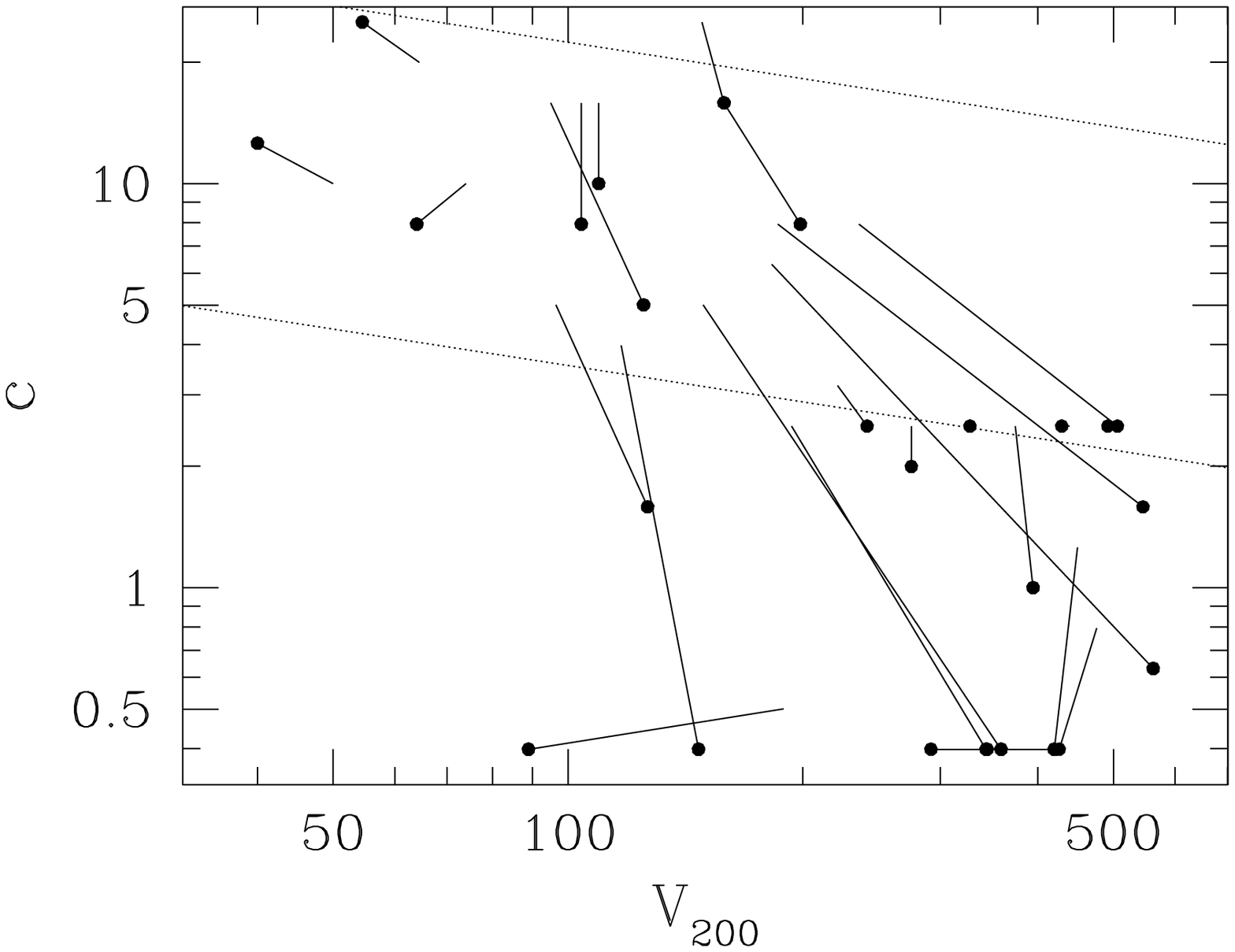}
\hfill\ \ 
\epsfysize=4.5cm
\epsfbox[44 169 528 539]{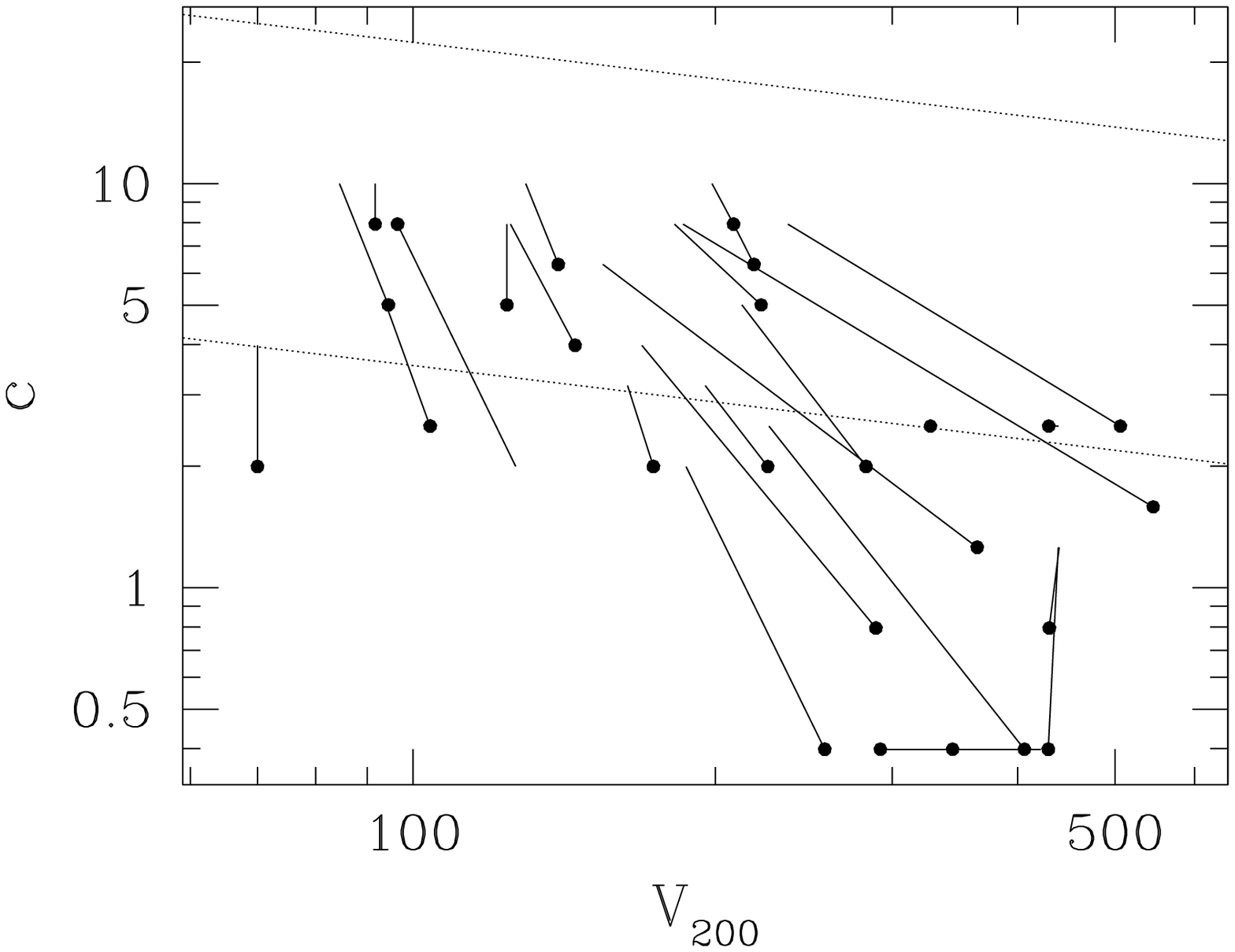}
}
\caption{
{\bf Left (a):} 
The concentration index $c$ versus virial speed $V_{200}$ of the
fitted NFW halo after adiabatic contraction (filled circles) and
before contraction (other end of line). The dotted lines indicate the
typical range for these parameters in $\Lambda$CDM models.
{\bf Right (b):}
The same, but for fits to rotation curves that were extended by a
factor 5 in radius with the last measured rotation speed.
}
\end{figure}

In Fig.\,3a we plot the resulting parameters characterizing an NFW
profile, $c$ and $V_{200}$. These parameters have have strongly
correlated errors (see Kassin, these proceedings; Dutton et al.~2003)
and are often poorly determined even with the stellar mass fixed (all
points at $c=0.4$ are basically undetermined fits). We plot only
galaxies with rotation curves extending to at least 3 effective mass
radii (about 5 disk scalelengths for pure exponential disks). Before
adiabatic contraction, most galaxies lie in the parameter space
expected by $\Lambda$CDM models (Bullock et al.~2001), but after
contraction many galaxies have too small $c$ for their $V_{200}$,
especially at the high mass end. These galaxies have $V_{200}$ values
that are higher than their highest measured speeds, meaning that the
model predicts that they have rising rotation curves. Because this has
never been observed for massive galaxies and having data out to 5 disk
scalelengths is apparently not enough of a constraint, we show in
Fig.\,3b fits with the observed rotation extended in radius by a
factor of 5 with the last measured speed. Even though this does bring
a few more galaxies into an acceptable range, there is still a
considerable number of galaxies with very high $V_{200}$ forced by the
fit of the central regions. We could reduced the problems caused by
adiabatic contraction by lowering our color-$M/L$ normalization, but
that would make all galaxies substantial sub-maximal, even the high
surface brightness ones.

\section{Specific Angular Momentum}

Galaxies are thought to build up their angular momentum through tidal
torques (Peebles 1969) or by the merging process during galaxy
assembly (e.g., Vitvitska 2002). N-body simulations of DM halo
formation by Bullock et al.~(2001) showed that the specific angular
momentum distribution in these DM halos follows a well defined shape.
%even though the gas in these halos may have a
%somewhat different distribution (van den Bosch et al.~2002). 
In semi-analytic models it is often assumed that the scale size of a
galaxy is set by detailed conservation of the original DM halo angular
momentum when the baryons collapse (e.g., Cole et al.~2000; de Jong \&
Lacey 2001).  We calculated the baryonic angular momentum
distributions for our best fit models and compared those to the
predicted distribution of $\Lambda$CDM models. Like van den Bosch et
al.~(2001) for dwarf galaxies, we find that all galaxies have angular
momentum distributions that are clearly different from the
models. This suggests that specific angular momentum is not conserved
when the gas collapses, which is not surprising given the
results of N-body/hydrodynamical modeling of galaxy formation by
e.g., Navarro \& Steinmetz (2000).

\section{Scaling the H{\small I} Contribution}

It has often been noted that the shapes of the H{\small I} and DM
rotation curves of galaxies are very similar and that a considerable
fraction of the DM in disk galaxies could be made of very cold
molecular gas distributed similarly to the H{\small I} (see
contributions by Combes, Pfenniger and Allen in these
proceedings). Hoekstra et al.~(2001) studied this phenomenon and found
that most of their 26 galaxies were well fitted with a scaled up
version of the H{\small I} mass distribution when their stellar $M/L$
value was left as a free parameter. Even with $M/L$ values determined
by their colors many of our galaxies can be fitted with such a scaled
H{\small I} model, but the needed H{\small I} scaling factors vary
greatly among galaxies. A secondary tuning parameter will be needed
for the scaling factor.

\section{Conclusions}

Stellar population synthesis models predict a strong color-$M/L$
relation. We use this correlation to model rotation curves of
galaxies and determine their DM distribution. Some of the main conclusions
of our investigation are:\\
- There are large degeneracies in the fitted NFW parameters, even with the
use of stellar $M/L$ constraints.\\
- With the chosen color-$M/L$ normalization, adiabatically contracted
halos not only over-predict central rotation
speeds of dwarf galaxies, but also those of high surface brightness galaxies.\\
- The angular momentum distribution in the observed baryons
differs from $\Lambda$CDM DM halo predictions, suggesting that
detailed angular momentum is not conserved during the galaxy assembly
process.\\
- H{\small I} rotation curves cannot consistently be scaled
to produce the DM signature.

%\acknowledgements


\begin{references}

\reference Bell, E.~F., \& de Jong, R.~S. 2001, ApJ, 550, 212
\reference Bell, E.~F., McIntosh, D.~H., Katz, N., \& Weinberg, M.~D.\ 2003, astro-ph/0302543
\reference Blumenthal, G.~R., Faber, S.M., Flores, R., \& Primack, J.R.\ 1986, \apj, 301, 27
\reference Bosma, A.\ 1981, AJ, 86, 1971
\reference Bullock, J.~S., Kolatt, T.~S., Sigad, Y., Somerville, R.~S., Kravtsov, A.~V., Klypin, A.~A., Primack, J.~R., \& Dekel, A.\ 2001, \mnras, 321, 559
%\reference Casertano, S.\ 1983, \mnras, 203, 735 
\reference Cole, S., Lacey, C.~G., Baugh, C.~M., \& Frenk, C.~S.\ 2000, \mnras, 319, 168
\reference Courteau, S.~\& Rix, H.\ 1999, \apj, 513, 561 
\reference de Jong, R.~S.~\& Lacey, C.\ 2000, \apj, 545, 781
\reference Dutton, A.A., Courteau, S., Carignan, C., \& de Jong, R.S. 2003, astro-ph/0310001
\reference Freeman, K.~C., 1970, ApJ, 160, 811
\reference Hoekstra, H., van Albada, T.~S., \& Sancisi, R.\ 2001, \mnras, 323, 453
\reference Navarro, J.~F.~\& Steinmetz, M.\ 2000, \apj, 538, 477 
\reference Navarro, J.~F., Frenk, C.~S., \& White, S.~D.~M.\ 1997, ApJ, 490, 493
\reference Peebles, P.~J.~E.\ 1969, \apj, 155, 393 
\reference Rubin, V.~C., Thonnard, N., \& Ford, W.~K.\ 1978, \apjl, 225, L107
%\reference Salpeter, E.~E.\ 1955, \apj, 121, 161
\reference van Albada, T.S., Bahcall, J.N., Begeman, K., \& Sancisi,
R.\ 1985, ApJ, 295, 30
\reference van den Bosch, F.~C.~\& Swaters, R.~A.\ 2001, \mnras, 325, 1017
%\reference van den Bosch, F.~C., Abel, T., Croft, R.~A.~C., Hernquist, L., \& White, S.D.M.\ 2002, \apj, 576, 21
\reference Verheijen, M.~A.~W.\ 1997, PhD thesis, Univ. of Groningen
\reference Vitvitska, M., et al. 2002, \apj, 581, 799
%\reference Vitvitska, M., Klypin, A.~A., Kravtsov, A.~V., Wechsler, R.~H., Primack, J.~R., \& Bullock, J.~S.\ 2002, \apj, 581, 799
\end{references}
\end{document}